\documentclass[aps,preprint]{revtex4}%
\usepackage{amsfonts}
\usepackage{amsmath}
\usepackage{amssymb}
\usepackage{graphicx}%
\providecommand{\U}[1]{\protect\rule{.1in}{.1in}}
\newtheorem{theorem}{Theorem}

\newtheorem{proposition}[theorem]{Proposition}

\begin{document}
\preprint{ }
\title{Physical Metric and the Nature of Gravity}
\author{Yukio Tomozawa}
\email{tomozawa@umich.edu}
\affiliation{Michigan Center for Theoretical Physics and Randall Laboratory of Physics,
University of Michigan, Ann Arbor, MI. 48109-1120, USA}
\author{}
\date{\today}

\begin{abstract}
A physical metric is defined as one which gives a measurable speed of light
throughout the whole space time continuum. It will be shown that a metric
which satisfies the condition that speed of light on the spherical direction
is that in a vacuum gives a correct result. All the metric functions thus
obtained are positive definite and exhibits a repulsive force at short
distances. The horizon in the sense of vanishing of the speed of light still
exists in the radial direction. It is located at $3\sqrt{3}r_{s}/2=2.60$
$r_{s}$, where $r_{s}=2GM/c^{2}$ is the Schwarzschild radius. This radius
corresponds to the size of a black hole, as well as the photon sphere radius.
The metric can be used to calculate general relativistic predictions in higher
order for any process.

\end{abstract}

\maketitle

\section{\label{sec:level1}Introduction}

The Schwartzschild metric is the exact solution of the Einstein equation of
general relativity\cite{mtw},\cite{weinberg}. It is a metric for the
spherically symmetric and static (SSS) system. However, since the speed of
light in a spherical direction inside the horizon of the Schwarzschild metric
is imaginary, it is not a physical metric. This characteristic, of course, is
changed if the role of r and t is awitched inside the horizon. The speed of
light becomes positive definite even inside the horizon. Then, the meaning of
static is lost, i. e. the metric inside the horizon is non-static. In this
article, the author constructs a physical metric by a coordinate
transformation of the Schwarzschild metric, maintaining the characteristic of
static nature and discusses the nature of gravity in the obtained physical
metric. In a forthcoming article, the author discusses the difference between
the Schwarzschild metric and the physical metric for the prediction of time
delay experiment of Shapiro et.al\cite{timedelay}.

\section{Asymptotic form for the physical metric}

The physical metric is expressed as
\begin{equation}
ds^{2}=e^{\nu(r)}dt^{2}-e^{\lambda(r)}dr^{2}-e^{\mu(r)}r^{2}(d\theta^{2}%
+\sin^{2}\theta d\phi^{2}), \label{eq0}%
\end{equation}
for a mass point $M$ in a spherically symmetric and static (hereafter refered
as SSS) metric. From the fact that the transformation, $r^{\prime}%
=re^{\mu(r)/2}$, leads to the Schwarzschild metric\cite{smoriginal}, one can
deduce the expression for the metric,%
\begin{equation}
e^{\nu(r)}=1-(r_{s}/r)e^{-\mu(r)/2}, \label{metric1}%
\end{equation}%
\begin{equation}
e^{\lambda(r)}=(\frac{d}{dr}(re^{\mu(r)/2}))^{2}/(1-(r_{s}/r)e^{-\mu(r)/2}),
\label{metric2}%
\end{equation}
where $r_{s}=2GM/c^{2}$ is the Schwarzschild radius. An asymptotic expansion
for the metric functions can be obtained from Eq.(\ref{metric1}) and
Eq.(\ref{metric2}), yielding%
\begin{equation}
e^{\nu(r)}=\sum_{n=0}^{\infty}a_{n}(r_{s}/r)^{n},\text{ }e^{\lambda(r)}%
=\sum_{n=0}^{\infty}b_{n}(r_{s}/r)^{n},\text{ }and\text{\ \ }e^{\mu(r)}%
=\sum_{n=0}^{\infty}c_{n}(r_{s}/r)^{n}, \label{eqasympt}%
\end{equation}
where%
\begin{align}
a_{0}  &  =b_{0}=c_{0}=1,\label{eq3}\\
-a_{1}  &  =b_{1}=1\text{ \ \ }and\label{eq4}\\
a_{2}  &  =c_{1}/2,\text{ \ }b_{2}=1-c_{1}/2+c_{1}^{2}/4-c_{2},\text{ \ }etc.
\label{eq5}%
\end{align}
It is obvious that $a_{n+1\text{ }}$and $b_{n}$ can be expressed as functions
of $c_{n}$, $c_{n-1}$ $\ldots$, $c_{1}$.

\section{Construction of the physical metric in the asymmptotic region}

Since the trouble of the Schwarzschild metric lies in the speed of light on a
spherical direction inside the horizon, one can eliminate this trouble by
requiring the following ansatz.

\begin{proposition}
The speed of light in the angular direction in the SSS metric is the same as
that of vacuum.
\end{proposition}

In other words, we require%
\begin{equation}
e^{\nu(r)}=e^{\mu(r)}=\omega
\end{equation}
This ansatz implies that although gravity deforms the geometry of space-time,
speed of light perpendicular to the gravity will not be affected.

Then one gets the equation for the asymptotic solution,%
\begin{equation}
e^{\nu(r)}=1-(r_{s}/r)e^{-\mu(r)/2}=e^{\mu(r)}=\omega.
\end{equation}
Then one has%
\begin{equation}
r_{s}/r=e^{\mu(r)/2}(1-e^{\mu(r)})=\omega^{1/2}(1-\omega), \label{eq9}%
\end{equation}
or%
\begin{equation}
(r_{s}/r)^{2}=\omega(1-\omega)^{2}. \label{eq10}%
\end{equation}
Differentiating Eq.(\ref{eq10}), one gets%
\begin{equation}
r\frac{d\omega}{dr}=\frac{2r_{s}^{2}/r^{2}}{(1-\omega)(3\omega-1)}%
=\frac{2\omega(1-\omega)}{(3\omega-1)}. \label{eq10.5}%
\end{equation}
From Eq.(\ref{metric2}), the metric function in the radial direction can be
calculated%
\begin{equation}
e^{\lambda(r)}=(\frac{d}{dr}(r\omega^{1/2}))^{2}/\omega=(\omega^{1/2}%
+\omega^{-1/2}r\frac{d\omega}{dr}/2)^{2}/\omega=(\frac{2\omega}{3\omega
-1})^{2}. \label{eq11.1}%
\end{equation}

\bigskip%
\begin{figure}[ptb]%
\centering
\includegraphics[
height=6.474in,
width=8.0739in
]%
{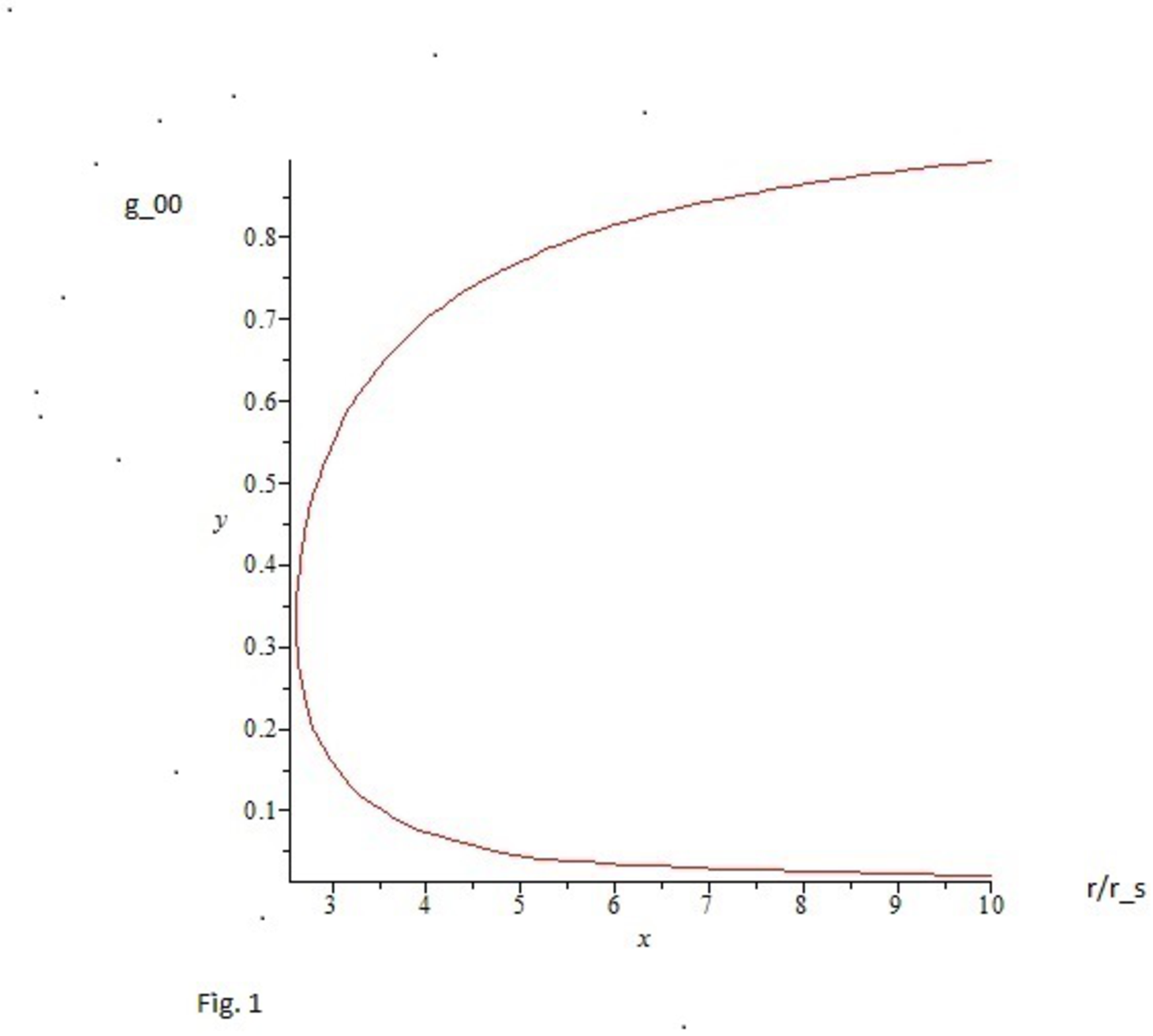}%
\end{figure}

From Eq.(\ref{eq9}) or Eq.(\ref{eq10}) and Fig. 1, it is clear that one covers
the range of
\begin{equation}
1>\omega>1/3
\end{equation}
and%
\begin{equation}
\infty>r/r_{s}>3\sqrt{3}/2.
\end{equation}

In order to cover the range of%
\begin{equation}
r/r_{s}<3\sqrt{3}/2,
\end{equation}
one has to use non-asymptotic solution of the Schwarzschild solution. From
Appendix, such a solution is given in the next section.

The asymptotic expansion of the metric functions can be calculated from
Eq.(\ref{eq10}) and Eq.(\ref{eq11.1}) as%
\begin{equation}
\omega=e^{\nu(r)}=e^{\mu(r)}=1-(r_{s}/r)-\frac{1}{2}(r_{s}/r)^{2}-\frac{5}%
{8}(r_{s}/r)^{3}-(r_{s}/r)^{4}-.....
\end{equation}
and%
\begin{equation}
e^{\lambda(r)}=1+(r_{s}/r)+\frac{9}{4}(r_{s}/r)^{2}+\frac{43}{8}(r_{s}%
/r)^{3}+\frac{211}{16}(r_{s}/r)^{4}+......
\end{equation}
where Eq.(\ref{eq11.1}) has been used. Successive expansion yields a
determination of all the parameters, $c_{n}$, for the physical metric. These
are useful for testing obervational data in higher order in gravity.
Alternatively, the inverse function of Eq.(\ref{eq9}) or Eq.(\ref{eq10}) may
be used.

\section{The physical metric in the whole region}

The Schwarzschild solution for non-asymptotic region (See Appendix section)
can be written as%
\begin{equation}
e^{\lambda(r)}=(1+\frac{Dr_{s}}{r^{\prime}})^{=1}%
\end{equation}
and%
\begin{equation}
e^{\nu(r)}=\frac{1}{A}(1+\frac{Dr_{s}}{r^{\prime}}),
\end{equation}
where A and D are non-dimensional constants. The metric functions for the
physical metric in the region%
\begin{equation}
r/r_{s}<3\sqrt{3}/2
\end{equation}
are expressed as%
\begin{equation}
\omega=e^{\nu(r)}=\frac{1}{A}(1+D(r_{s}/r)e^{-\mu(r)/2})=e^{\mu(r)}%
\end{equation}
and%
\begin{equation}
e^{\lambda(r)}=(\frac{d}{dr}(r\omega^{1/2}))^{2}/A\omega, \label{eq11.5}%
\end{equation}
and hence%
\begin{equation}
D(\frac{r_{s}}{r})=\omega^{1/2}(A\omega-1) \label{eq12}%
\end{equation}
or%
\begin{equation}
(\frac{Dr_{s}}{r})^{2}=\omega(A\omega-1)^{2}%
\end{equation}
Differentiating Eq.(\ref{eq12}), one gets%
\begin{equation}
r\frac{d\omega}{dr}=-2\omega\frac{A\omega-1}{3A\omega-1}%
\end{equation}
and%
\begin{equation}
e^{\lambda(r)}=(\omega^{1/2}+r\frac{d\omega}{dr}/2\omega^{1/2})^{2}%
/A\omega=A(\frac{2\omega}{3A\omega-1})^{2} \label{eq12.5}%
\end{equation}

Imposing the continuity of the asymptotic expression, Eq.(\ref{eq9}) and the
non-asymptotic expression, Eq.(\ref{eq12}) at%
\begin{equation}
(r/r_{s},\omega)=(3\sqrt{3}/2,1/3) \label{eq12.0}%
\end{equation}
one gets%
\begin{equation}
A=2D+3.
\end{equation}
One splits the parameter space in the following 3 Regions.

I) $A>3$ and $D>0,$

II) $3>A>0$ and $0>D>-\frac{3}{2}$

III) $0>A$ and $-\frac{3}{2}>D$.

In Region I, the distance r can be reached at zero when $\omega$ reaches
$\infty$, as%
\begin{equation}
\omega=(\frac{Dr_{s}}{Ar})^{2/3} \label{eq13}%
\end{equation}
In Region II, the distance r cannot reach at zero value (the origin). In
Region III, the distance r can be reached at zero, as in Eq.(\ref{eq13}).
However, the radial metric function, $e^{\lambda(r)}$, becomes negative as is
seen from Eq.(\ref{eq11.5}) or Eq.(\ref{eq12.5}). This means that the radial
velocity of light becomes imaginary. Therefore the most natural choice for the
parameter space is the Region I, where all metric functions are positive
definite. This is a characteristic of the physical metric.

\section{The nature of the physical metric}

Summarizing the previous sections, one obtains the physical metric%
\begin{equation}
e^{\nu(r)}=e^{\mu(r)}=\omega
\end{equation}
in the following manner. For the asymptotic region%
\begin{equation}
r/r_{s}\geq3\sqrt{3}/2,
\end{equation}
one has%
\begin{equation}
r_{s}/r=\omega^{1/2}(1-\omega) \label{seq1}%
\end{equation}
and%
\begin{equation}
e^{\lambda(r)}=(\frac{2\omega}{3\omega-1})^{2}. \label{seq1.5}%
\end{equation}
The range of $\omega$ is restricted to%
\begin{equation}
\omega\geq1/3.
\end{equation}
For the non-asymptotic region%
\begin{equation}
3\sqrt{3}/2\geq r/r_{s}>0,
\end{equation}
one has%
\begin{equation}
D(\frac{r_{s}}{r})=\omega^{1/2}(A\omega-1) \label{seq2}%
\end{equation}
and%
\begin{equation}
e^{\lambda(r)}=A(\frac{2\omega}{3A\omega-1})^{2}.
\end{equation}
The continuity of Eq.(\ref{seq1}) and Eq.(\ref{seq2}) at%
\begin{equation}
(r_{s}/r,\omega)=(3\sqrt{3}/2,1/3) \label{eqmeet}%
\end{equation}
requires%
\begin{equation}
A=D/2+3.
\end{equation}
Choosing the range of the parameter space to be%
\begin{equation}
D>0,\text{ \ \ and \ \ }A>3,
\end{equation}
all the metric functions are positive definite, which guarantees the condition
for the physical metric, having the definite value for speed of light
throughout the whole space-time continuum. \ 

At the origin, the metric function, $\omega=g_{00}(r)$, diverges as
\begin{equation}
\omega=(\frac{Dr_{s}}{Ar})^{2/3}.
\end{equation}
From Eq.(\ref{seq1.5}), one concludes that $e^{\lambda(r)}=g_{11}(r)$ becomes
$\infty$ at the edge point of the asymptotic region, Eq.(\ref{eqmeet}). Hence,
at $r_{s}/r=3\sqrt{3}/2$, the radial speed of light vanishes. One may call
this point a horizen, although the characteristics are very different from
that of the Schwartzscield metric. The speed of light in the spherical
direction is that of vacuum. Since all metric are positive definite in the
whole region, speed of light is well defined throughout the whole space time.
The magnitude of the horizon is $(3\sqrt{3}/2)$ $r_{s}=2.60$ $r_{s}$, i.e.,
2.6 times bigger than that of the Schwartzschild radius. Below Fig. 2 showes
the picture of $g_{00}(r)=$ $e^{\nu(r)}=\omega$ as a function of $r/r_{s}$,
namely the picture of the gravitational potential with the shift of the y axis
and a scale factor of 2.

\bigskip%
\begin{figure}[ptb]%
\centering
\includegraphics[
height=6.474in,
width=8.0739in
]%
{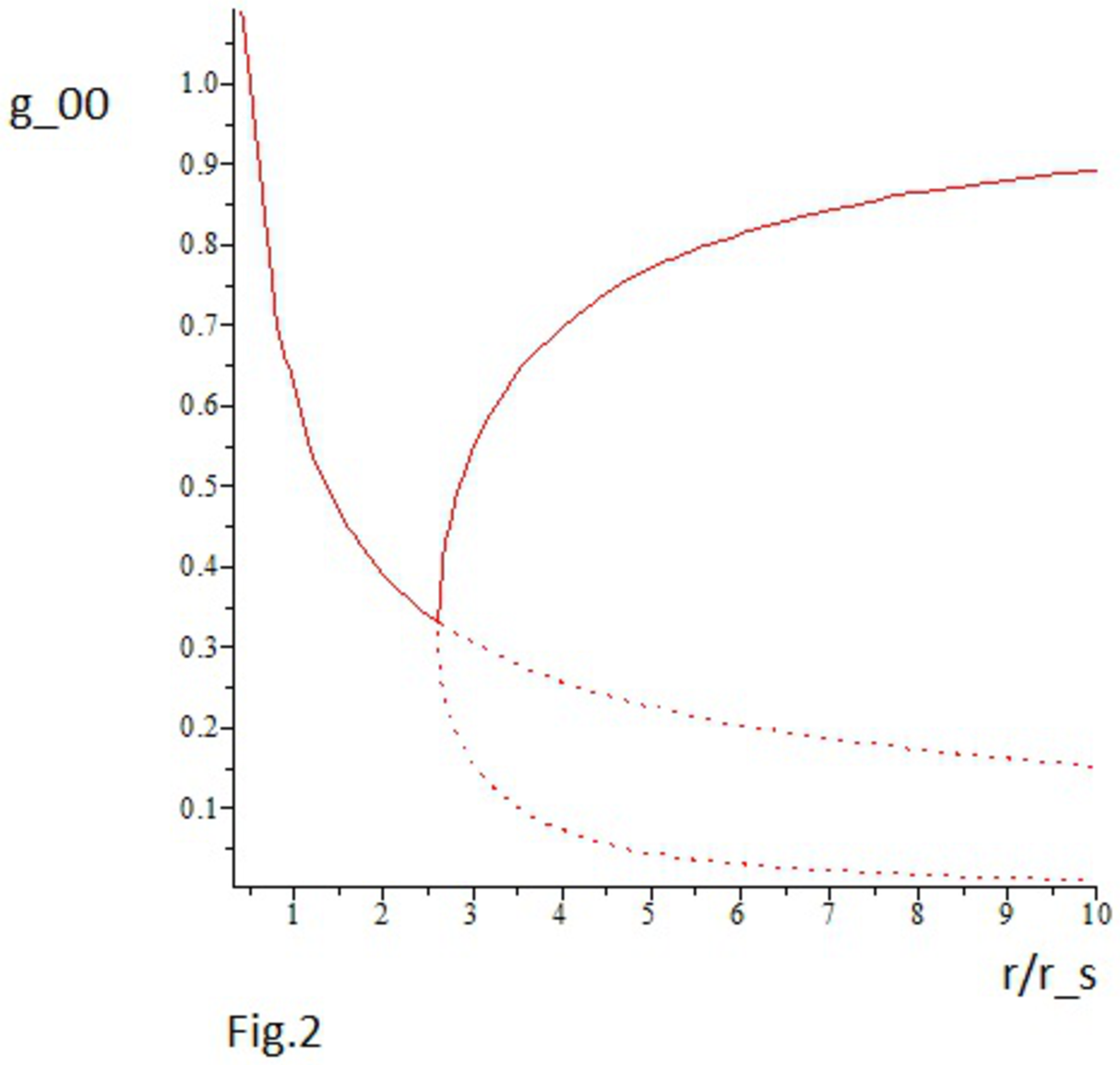}%
\end{figure}

\bigskip

The crossing point of the two curves is%
\begin{equation}
(r/r_{s},\omega)=(3\sqrt{3}/2,1/3).
\end{equation}
The radial metric function, $g_{11}=e^{\lambda(r)},$ is inevitably
discontinuous at this point. This discontinuity allows the passing of all
particles through the horizon, in and out. More importantly, the gravitational
potential inside the horizon is repulsive. This property could change the
nature of gravity, black holes, cosmic ray production as well as the nature of
cosmology. I will discuss these problems in the forthcoming articles.

\section{Summary}

The author has constructed a physical metric for which the speed of light is
well defined throughout all space time continuum. The constructed metric shows
that gravity is repulsive inside the horizon and the size of black holes is 3
times bigger than the Schwarzschild radius. The latter can be tested by
observation in the near future\cite{haystack}. The observational effects of
the physical metric will be discussed in forthcoming articles.

\section{Appendix. The Schwarzschild solution}

Setting%
\begin{equation}
e^{\mu(r)}=1,
\end{equation}
in Eq.(\ref{eq0}), and using the Maple program the Einstein equation reads%
\begin{equation}
-r\lambda^{\prime}(r)-e^{\lambda(r)}+1=0, \label{eq01}%
\end{equation}

\begin{equation}
-r\nu^{\prime}(r)+e^{\lambda(r)}-1=0 \label{eq02}%
\end{equation}
and%
\begin{equation}
2\nu^{\prime}(r)-2\lambda^{\prime}(r)+2r\nu^{\prime\prime}(r)+r\nu^{\prime
}(r)^{2}-r\nu^{\prime}(r)\lambda^{\prime}(r)=0. \label{eq03}%
\end{equation}
From the sum of Eq.(\ref{eq01}) and Eq.(\ref{eq02}), one gets%
\begin{equation}
\nu^{\prime}(r)+\lambda^{\prime}(r)=0. \label{eq04}%
\end{equation}
Using this relation, Eq.(\ref{eq03}) becomes%
\begin{equation}
-r\lambda^{\prime\prime}(r)+r\lambda^{\prime}(r)^{2}-2\lambda^{\prime}(r)=0
\end{equation}
or equivalently%
\begin{equation}
e^{\lambda(r)}(re^{-\lambda(r)})^{\prime\prime}=0. \label{eq05}%
\end{equation}
On the other hand, Eq.(\ref{eq01}) can be written as%
\begin{equation}
(re^{-\lambda(r)})^{\prime}=1, \label{eq06}%
\end{equation}
which solution is%
\begin{equation}
e^{-\lambda(r)}=1+\frac{B}{r},
\end{equation}
and Eq.(\ref{eq05}) is satisfied, where B is an integration constant. The
solution of Eq.(\ref{eq04}) reads%
\begin{equation}
e^{\nu(r)}=\frac{1}{A}(1+\frac{B}{r}).
\end{equation}
The asymptotic solution with the boundary condition is given by%
\begin{equation}
A=1,\text{ \ \ }B=-r_{s}.
\end{equation}
On the other hand, the non-asymptotic solution is given by%
\begin{equation}
B=Dr_{s},\text{ \ \ }A\text{ }arbiray.
\end{equation}
where A and D are nondimensional integration constants.

\begin{acknowledgments}
It is a great pleasure to thank Peter K. Tomozawa for reading the manuscript.
\end{acknowledgments}

Figure captions

Fig. 1 The metric function, $g_{00}(r)$, in the asymptotic region in the SSS
physical metric.

Fig. 2 The metric function, $g_{00}(r)$, as a funtion of r/r$_{s}$ in the SSS
physical metric.

\end{document}